\documentclass[useAMS]{mn2e}
\usepackage{times}

\def\be{\begin{equation}}
\def\ee{\end{equation}}
\def\ba{\begin{eqnarray}}
\def\ea{\end{eqnarray}}

\title[Starbursts] {\Large \bf Ultraluminous Starbursts  from SMBH-induced outflows }
\author[Silk ]
{Joseph Silk \\Astrophysics, University of Oxford \\
Denys Wilkinson Building, Keble Road, OX1 3RH Oxford, UK}

\begin{document}

\date{Draft version \today}

\pagerange{\pageref{firstpage}--\pageref{lastpage}} \pubyear{2005}

\maketitle

\label{firstpage}

 \begin{abstract}
I argue that there are two modes of global star formation.  Disks and smaller
spheroids form
stars relatively inefficiently  as a consequence of supernova-triggered
negative feedback via a sequence of ministarbursts (S mode), whereas 
massive spheroids formed rapidly with high
efficiency via the impact of AGN jet-triggered positive feedback (J mode)
that generates and enhances ultraluminous starbursts. Supermassive black hole
growth by accretion is favoured in the gas-rich protospheroid
environment as mergers build up the mass of the host galaxy and
provide a centrally concentrated gas supply.   Quasi-spherical outflows
arise and provide the source of porosity as the energetic jets from
the accreting central SMBH are isotropised by the inhomogeneous
interstellar medium in the protospheroid  core. Super-Eddington outflows occur
and help generate  both the SMBH  at high redshift and the strong positive feedback
on protospheroid star formation that
occurs as dense interstellar
clouds are overpressured and collapse. SMBH form before 
the bulk of spheroid stars,
and the correlation between spheroid velocity dispersion and
supermassive black hole mass arises as AGN-triggered outflows limit
the gas reservoir for spheroid star formation.
The super-Eddington phase plausibly triggers a top-heavy IMF in the 
region of influence of the SMBH.
The Compton-cooled Eddington-limited outflow phase  results
in a spheroid core  whose phase space density scales as the inverse 5/2 power of the core mass,
and whose mass scales as the 3/2 power of SMBH mass. This latter  scaling suggests that SMBH growth 
(and hence spheroid formation) is anti-hierarchical.   

\end{abstract}

\def\simlt{\lower.5ex\hbox{$\; \buildrel < \over \sim \;$}}
\def\simgt{\lower.5ex\hbox{$\; \buildrel > \over \sim \;$}}
\def\simpropto{\lower.2ex\hbox{$\; \buildrel \propto \over \sim \;$}}


\section{Introduction}
It has long been argued on the basis of galaxy colours that star
formation is bimodal. Disks are blue and have extended star formation
over a Hubble time. Ellipticals, S0s and bulges are red, and are
required to
have formed in a burst of star formation that lasted $\sim \rm
10^8 yr.$  Very different star formation efficiencies are inferrred,
and this requirement has motivated the idea that there might be two distinct star formation modes,
one possibly associated with a top-heavy initial mass function ({\it e.g.},
Larson 1986;  Sandage 1986).

In fact, the situation is not completely bimodal.
Modelling of spectral energy distributions requires the
star formation rate in disks, conveniently parametrised by
$b=SFR/<SFR>$, to vary from $\sim 0.1$ in early-type disks to $\sim
10$ in late-type disks.  Moreover, the past history of star formation
in the solar neighbourhood is not monotonic, showing signs of minibursts of star formation
(Rocha-Pinto et al 2000).
Also,  many nearby early-type galaxies, ellipticals and
S0s, show traces of ongoing or recent star formation when observed in
the FUV. The rates are low, and consistent with a very low efficiency, $b =0.01-0.001.$
 In contrast, high redshift observations of ULIRGs
at $z\sim 6 ,$ where 
 the high star                                            
formation rate is indicative of elliptical formation, 
require an extremely high efficiency of star 
formation.

Clearly, the concept of two modes of global star formation is likely
to be an oversimplification of the physical situation.
However for any specified gas initial supply, it is clear that star 
formation is intrinsically inefficient in disks and relatively efficient 
when ellipticals formed. 
The inefficiency of star formation in cold, gas-rich disks is reasonably well understood as
a consequence of disk gravitational instability and supernova-driven  
feedback  (Silk 1997, 2003; Efstathiou 2000).

Another requirement is outflow.
Galactic winds are required in order to account for the enrichment of the warm intergalactic medium
and intracluster gas.
Winds are  both predicted and observed in dwarfs, but are  more problematic for massive galaxies, 
at least in so far as the theory is concerned.
 Some evidence for substantial outflows from massive galaxies at high redshift is inferred
directly, for Lyman break galaxies at $z\sim 3$ that display offsets between stellar and (blue-shifted) interstellar spectral
lines as well as possible Mpc-scale cavities in the surrounding IGM
(Shapley et al. 2003, Adelberger et al. 2003),
and for ULIRGs where high velocity cold outflows are observed.
That the Lyman
break galaxies are indeed massive systems has recently been verified
(Adelberger et  al. 2004)
 by their clustering properties at $z\sim 2.$
In the case of our own galaxy, approximately half of the baryons initially
present most likely were ejected (Silk 2003). This presumably ends up in the 
IGM (both photo-ionised Lyman alpha forest and collisionally ionised WHIM),
which together contain about half of the baryons in the universe, enriched to 
approximately 10\% of the solar metallicity at the present epoch.
In summary, a plausible case can be made that
a mass in baryons of order that in stars must have been ejected in the early 
stages of galaxy formation.

In this note, I develop a complementary theory of efficient star formation in protospheroids that 
is  appropriate to ultraluminous starbursts and 
massive spheroid formation. The initial phase of 
black hole growth is associated with super-Eddington accretion: this both allows  rapid SMBH growth at high redshift
and efficient star formation feedback induced by super-Eddington outflows.
Other outcomes include the predictions of momentum-driven
winds  from ULIRGS with $v_w \simpropto \dot
M_\ast^{1/2},$  
a spheroid core phase space density  that scales with core mass $M_c$ as $\rho_c/\sigma_g^3\propto
M_c^{-5/2},$ 
and  a supermassive black hole mass that scales as $M_{BH}=f_g\frac{\sigma_T}{G^2 m_p}\sigma_g^4$
and $\propto M_c^{2/3}.$

Some of these ideas are not new. The notion that outflow from 
a quasar phase quenches both black hole growth and star formation 
in such a way as to account for the observed correlation between mass of the
SMBH and spheroid velocity dispersion  was
pioneered by Silk and Rees (1998) and implemented in  recent 
simulations of galaxy formation (Romano et al. 2002;
Granato et al. 2004; Di Matteo, Springel and Hernquist 2005). 
The case for a super-Eddington SMBH growth phase was made by Haiman (2004).
Triggering of star formation in radio lobes has been proposed to explain  the
alignments of optical continuum and emission along the radio axis
(De Young 1989; Rees 1989). Extensive  star formation  as well as generation
of intergalactic magnetic fields and metal enrichment has been argued to
occur during the peak of quasar activity ($1.5\simlt z \simlt 3$)
(Gopal-Krishna
and Wiita 2003; Gopal-Krishna,  Wiita  and Barai 2004). The general
connection between AGNs and spheroid formation was proposed and developed
 by Chokshi
(1997).
The novel idea put forward here is to incorporate many of these ideas into 
the modern view of hierarchical galaxy formation. Star formation time-scales
are the key to understanding how disks and spheroids are formed. 
Supernova-induced feedback controls disk and dwarf galaxy formation 
by rendering star formation relatively slow and inefficient. 
 AGN outflows trigger star formation
on a short time-scale by overpressuring protogalactic clouds.  Hence they 
provide an efficient star formation mode
before the combined momentum input from both the outflows and SNe eventually
drives out the residual gas,  suppresses  SMBH growth by accretion, and terminates
the starburst.

\section{Disk mode}

I summarise here the key elements of a simple model for global star formation in
disks (Silk 2003).
A key parameter in the model 
is the filling factor $f_h$ of hot gas which can be expressed in terms of the
 porosity  $Q$ by $1-e^{-Q}.$ 
The star formation rate in a disk containing $M_{gas}$ in cold
gas and with gas fraction $f_g$ is $\dot M_\ast=Q\epsilon
M_{gas}\Omega=\alpha_S f_g v_{circ}^3/G,$ where
$Q$ is the porosity, $\epsilon=(\sigma_g /\sigma_f)^{2.7}$, and
$\alpha_S =Q\epsilon=\sigma_g v_{cool} m_{SN}/E_{SN}.$ Here 
$\Omega$ is the disk rotation rate, $v_{circ}$ is the maximum rotation
velocity, $\sigma_g$ is the gas velocity dispersion,
$E_{SN}$ is the
kinetic energy of a supernova, $m_{SN}$ is the mass in forming stars
required to produce a Type II supernova
(approximately 200M$_\odot$ for a Kroupa IMF). Also
 $v_{cool}\approx 400\, \rm km\, s^{-1}$ is the shell velocity at which
strong radiative energy losses set in  and
$\sigma_f\approx 20 \rm km \, s^{-1}\left(E_{SN}/10^{51}\rm erg\right)^{0.6}
\left(\rm 200M_\odot/m_{SN}\right)^{0.4}$
 is a fiducial velocity. The latter is likely to be a lower bound   due to allowance for
more realistic physics such as  enhancements of porosity due {\it e.g.}
to Rayleigh-Taylor instabilities and deviations from spherical symmetry.

The  basic idea  is that supernovae explosively blow hot bubbles into
the protogalactic interstellar medium. 
The bubbles decelerate by sweeping up shells
of cold gas and eventually break up when the expansion is halted by
the ambient pressure.  If the rate of bubble formation is sufficiently
high, the bubbles overlap and a multi-phase medium develops of hot
shell-shocked gas in which dense cold shell fragments are embedded. If
the hot gas permeates through the cold gas scale height, galactic
fountains, chimneys  and  outflows will result. These phenomena, 
in what I refer to as the S mode of star formation, 
result in  negative feedback as a consequence of the supernova input. It is for this reason that the gas supply 
for star formation is limited but long-lived. 

This simple analytic model appears to incorporate at least some of the
crucial physics and has successfully been tested against simulations
of a kiloparsec cube of the interstellar medium 
(Slyz et al. 2005).
One can reproduce the
extreme star formation rate in a starburst as well as the more
quiescent low efficiency star formation that persists for up to a
Hubble time. For a typical gas turbulent velocity (the relative
motions of cold clouds), the star formation efficiency $\alpha_S= Q\epsilon$
is 
$$ \approx 0.02
\left(\frac{\sigma_{gas}}{10 \, {\rm km s^{-1}}}\right)
\left(\frac{v_c}{400 \, {\rm km s^{-1}}}\right)
\left(\frac{m_{SN}}{200 {\rm M_\odot}}\right)
\left(\frac{10^{51}{\rm ergs}}{E_{SN}}\right).$$
The observed mean value is 0.017 (Kennicutt 1998).

Self-regulation enters in the following way. In disks, the feedback is initially positive but  becomes negative once 
gas can vent
out in fountains. 
Negative feedback and low efficiency  is essential for understanding gas-rich disk longevity.
$Q$ is expected to be of order unity, as indeed
self-regulation plausibly requires.  In starbursts, star formation is
concentrated and more efficient.  The turbulent momentum driving is
larger ($\alpha_S \propto \sigma_g$) whereas the porosity is low,
because the increase in the porosity efficiency parameter $(\epsilon\propto \sigma_g^{2.7})$
overwhelms the increase in $\alpha_S .$ At low porosity, the 
initial feedback is positive.

Local starbursts can be modelled in detail.
Incorporation of a Schmidt-Kennicutt star formation law relating the
star formation rate to local cold gas density fails to explain the
extended nature of the star formation observed in merging
galaxies 
(Schweizer 2004).
Incorporation of turbulence into the expression for star formation
provides a greatly improved model for star formation in the Mice, a nearby  pair of merging
galaxies (Barnes 2004).

One immediate success of such a turbulent star formation
model is a refined fit to modelling star formation in merging galaxies.
Star formation globally is interpreted as a series of ministarbursts.
The gas supply is regulated by infall of small gas-rich satellites.
The Milky Way is an interesting case study.
The high velocity clouds may be manifestations of such objects.
There is increasing
evidence for past minor mergers, a form of gas accretion,
 from combining chemical and dynamical
tracers of high velocity stars. Chemical signatures associated with
the disruption of the Arcturus stream and with $\omega$ Cen, and
kinematical modelling of the Sagittarius dwarf tidal stream, all point
towards a merging history which leaves behind relic stellar tracers
(Navarro, Helmi and Freeman 2004; Helmi 2004).
Such merging events would have 
injected a sporadic supply of
gas that temporarily reinvigorated star formation. 
Evidence for such ``ministarbursts" is seen in the
local history of star formation as traced by chromospheric dating of
nearby stars (Rocha-Pinto et al. 2000)
Even  on the scale of individual  star-forming clouds
and young stellar associations, there is empirical
evidence that star formation is accelerating (Palla and Stahler 2000),
and this has been interpreted as evidence for ministarbursts (Silk 2004).

Perhaps one of the strongest arguments for a ministarburst history in
galaxies of stellar mass up to $\sim 10^{10}\rm M_\odot$ comes from
recent studies of luminous infrared galaxies at $z>0.4$ (Hammer et al 2005)
and of the mass-metallicity relation in the local universe
(Tremonti et al. 2004).
The high frequency of so-called LIRGs argues for an
episodic or bursty star formation history in intermediate mass
galaxies. Moreover, evidence for metal loss via winds is inferred not
 just for dwarfs but for moderately massive galaxies. Quiescent disk
star formation cannot drive outflows from such galaxies. A series of
blow-out events associated with ministarbursts may suffice. 
Accumulating evidence from studies of the local universe suggests that
nearby starbursts typically occur in sub-L$_\ast$ galaxies and have
outflows of order the star formation rate, at least for the handful of
well-studied examples, {\it} e.g. 
Martin (2002).

The  outflow rate can be estimated to be   the
product of the supernova rate, the hot gas filling factor, and the
mass loading factor (Silk 2003). The latter depends on such effects as
Kelvin-Helmholtz instabilities that entrain cold gas into the hot
phase. 

In a starburst, one can now express the rate of gas
outflow as $\dot
M_{outflow}=(1-e^{-Q}) \dot M_\ast f_L 
\sim Q^2 \epsilon\dot M_\ast.$
Rewriting this in terms of the gas  mass,
$\dot
M_{outflow}
\approx f_L\alpha_S^2\epsilon^{-1}M_{gas}\Omega .
$
Here $f_L$ is the mass-loading factor associated with entrainment of
cold gas and depends on the detailed structure of the interface
between hot and cold gas phases.  Kelvin-Helmholtz instabilities will
enhance mass loading. 
 Observationally, we know that for at least one
nearby starburst, $f_L \sim 10$ 
(Martin 2002).
One can attain $\dot
M_{outflow}\sim \dot M_\ast\sim 0.1 M_{gas}\Omega$ in dwarf starbursts, where 
the preceding estimates yield 
$\alpha_S\sim 0.1$, $Q\sim 0.1,$  and $\epsilon\sim 1.$
However for starbursts in deep potential wells, so that $\sigma_g\gg 30
\rm km s^{-1},$ one  has $\epsilon \gg 1, $ and 
outflows  induced by supernovae are suppressed.

Even inclusion of such effects as cloud disruption as well as more realistic
geometries and porosity boosting via inclusion of Kelvin-Helmholtz and Rayleigh-Taylor
instabilities are unlikely to raise $\sigma_f$ by more than a factor of 2. One
can gain perhaps another factor of 2-3 by allowance for hypernovae and
especially by inclusion of a top-heavy IMF. Galactic outflows cannot escape
from potentials where the escape velocity exceeds $\sim 100 \rm km\, s^{-1}$,
if supernova-driven. Nor is there  any effective  supernova-driven feedback
unless $\sigma_f \propto E_{SN}^{0.6}m_{SN}^{-0.4}$
is significantly  boosted  over the value expected for a standard IMF and supernova energy.
The porosity is simply too low.

\section{Spheroid mode}

It has long been conjectured that mergers of gas-rich galaxies trigger
violent star formation associated with spheroid formation. 
 However there is
no detailed modelling of the conversion of gas into stars and hence of
the star formation efficiency. The 
Schmidt-Kennicutt star formation law, in which star formation rate per
unit area is proportional to the product of gas surface density and
disk rotation rate, fits nearby disks and starbursts. 
Nevertheless, 
this approach must  be inadequate in the more extreme situations
associated with spheroid formation, and even, for example, with ULIRGs,
where extremely high star formation efficiencies are inferred.

Specifically, adoption of a universal star formation efficiency
 results in challenges to the comparison of semi-analytical galaxy
formation models with the observational data. The predicted colour and
age distributions, alpha element-to-iron abundance ($[\alpha/Fe]$) ratios
relative to solar, 
and infrared/submillimeter
galaxy counts are all in conflict with results from recent surveys
({\it e.g.},  Thomas et al.  2005).


It seems unlikely that supernovae  can account for the high star formation efficiency
inferred in the early universe, associated with spheroid formation.
Supernova-driven gas flows cannot fill  massive galaxy potential wells, and cannot cope with the
overcooling problem common to simulations, whereby an excess of massive
galaxies is  formed.
In  more extreme situations, supernovae 
cannot drive
  the massive
galactic outflows that are observed  especially
at high redshift. Here, one can also cite the enrichment of the IGM, and more directly,
the strong radio sources that are
centrally dominant cluster galaxies with extended Lyman alpha
emission (Reuland et al. 2003),
Lyman break galaxies, and ULIRGs.
Clearly, one needs an additional energy source.

A resolution to both the star formation efficiency and
wind dilemmas most
 likely comes from allowance for supermassive black hole (SMBH)-induced
 outflows.  Suppose that SMBH outflows provide an explanation for the
 correlation between black hole mass and spheroid velocity dispersion
 via their effect on gas retention and possibly also on star formation
 in the protogalaxy.  The correlation between central black hole mass
 and spheroid velocity dispersion suggests that the black hole
 forms contemporaneously with (Dietrich and Hamann 2004)
 or even before the stellar spheroid (Walter et al. 2004).

 There is indeed a
 natural coupling, since the SMBH
undergoes most of its growth   in the gas-rich phase and
 the SMBH outflow pressurises the gas.  Once the stars have formed,
 the outflow is irrelevant.

We also know from the observed local space density of SMBH
and the quasar luminosity function that, if quasars radiate near the
Eddington limit, radiatively inefficient accretion is largely
responsible for black hole growth (Soltan 1982; Yu and Tremaine 2002).
One infers
that black hole mergers play a subdominant role in growing SMBH,
unless the radiative efficiency is implausibly high and the amount of
accretion inferred is thereby reduced. However to account for the highest
redshift, ultraluminous quasars, formation of SMBH at $z\simgt 6-7$ in
the time available requires super-Eddington accretion rates ({\it c.f.} Haiman 2004)
as well as seed black holes of at least $\sim 1000\rm
M_\odot$ (Islam, Taylor and Silk 2004).

A bipolar gas outflow is likely to be an
 inevitable consequence of SMBH formation via disk accretion.
  The gas-rich protogalaxy provides the ideal accretion
environment for forming the SMBH. 
Overpressured cocoons, as inferred for high redshift radio galaxies,
 engulf and overpressure interstellar clouds (Begelman and Cioffi
 1989).  A broad jet that propagates through an inhomogeneous
 interstellar medium with a low dense cloud filling factor is
 disrupted and isotropises (Saxton et al. 2005)
to form an expanding cocoon.  The hot plasma cocoon overpressures cold
clouds and induces collapse within the central core of the forming
galaxy. Star formation is triggered coherently and rapidly in what I refer to as the J mode.
The feedback is positive, as there is insufficient time for the 
supernova-driven negative feedback to develop.
 The interaction of the outflow with the surrounding
protogalactic gas at first stimulates star formation on a short time-scale, $10^7$ yr or less. , but will
eventually expel much of the gas in a wind.  Evidence has been found
for jet-stimulated star formation up to $z\sim 5$ (Venemans et
al. 2004, 2005), followed in at least  one case by a starburst-driven superwind
(Zirm et al. 2005).

One may crudely describe this situation by modelling the late-time
cocoon-driven outflow as quasi-spherical, and 
use  a spherical shell approximation
to describe the swept-up protogalactic gas. Initially, following
Begelman and Cioffi (1989), the interaction of the pair of jets may be
modelled by introducing an overpressured and much larger cocoon, the ends of which
advance into the protogalactic gas at a speed $v_J$ determined by the jet
thrust independently of the ambient pressure, and which expands
laterally at a speed determined eventually  by pressure balance with the ambient
gas.  I first express the jet luminosity $L_J$ in terms of the critical luminosity needed to expel
all of the protogalactic gas.

 If outflow  limits the spheroid  star formation
by depleting the gas supply (Silk and Rees 1998, King 2003),
one  has a critical luminosity
$L_{cr}/v_J=GMM_g/r^2=f_g\sigma_g^4/ G,$
where $f_g$ is the initial gas fraction. 
If $L_{cr}/v_J=L_{Edd}/c
=4\pi G
\kappa^{-1}M_{BH},$ where $\kappa$ is the electron-scattering opacity,
it follows that $M_{BH}=f_g\sigma_g^4\kappa/4
\pi G^2.$ This is the observed correlation, in slope and
normalisation (Gebhardt et al. 2000; Ferrarese and Merritt 2000;
Onken et al. 2004).  The earlier short-lived phase of efficient star
formation is induced by outflow luminosities that are super-Eddington.
To estimate this effect,
first consider the jet/cocoon-protogalactic gas interaction.

In the initial outflow phase,
radial momentum flux balance sets $L_J/v_J$ equal to  
$\rho_a R_h^2 v_h^2,$
where $R_h$ is the hot spot radius. Transverse momentum flux balance
controls the lateral expansion of the cocoon, which is 
  wider than the hot spot 
 because of jet precession, 
and  sets $L_J/v_h$ equal to  
$\rho_a R_c^2 v_c^2,$
where $R_c$ is the cocoon width, $v_c$ is the cocoon expansion velocity, 
$\rho_a$ is the ambient gas density,
and
 $v_h>v_c$ is the hot spot velocity. I identify $v_h$ with the  wind velocity
$v_w$ and infer that 
$$\frac{L_J}{L_{cr}}=\frac{f_c}{f_g}
\left(\frac{v_c}{\sigma_g}\right)^2\left(\frac{v_w}{v_J}\right)^2
=\frac{f_c}{f_g}\left(\frac{v_w}{\sigma_g}\right)^2
\left(\frac{R_h}{R_c}\right)^2,
$$
where $f_c$ is the baryon compression factor in the core, enhanced due to
mergers. One expects in order of magnitude that
$(f_c/f_g)(R_h/R_c)^2\sim 1.$
Hence the early outflow   is plausibly super-Eddington 
by a factor of order $(v_w/v_c)^2$  until the cocoon becomes quasi-spherical
when its static pressure exceeds the time-averaged jet ram pressure. Expansion continues until 
the cocoon ceases to be overpressured once  $v_c\sim \sigma_g.$

Only a small fraction of the protogalactic gas reservoir is implicated in AGN
feeding, even if this occurs at the maximum (Bondi) accretion rate.
Super-Eddington outflow  of course requires
super-Eddington  accretion which is plausibly associated with an
Eddington luminosity-limited luminous phase   implicated in the need to
generate massive SMBH by $z\sim 6$ ({\it c.f.} Volonteri and Rees 2005).
As the hierarchy develops, gas-rich  mergers replenish the gas supply.
A time-scale of $10^6-10^7$ yr for the super-Eddington phase would
more than suffice to provide the accelerated triggering of star formation associated with the J mode.

Once the jet is quenched, the
super-Eddington phase is followed by a longer duration period of
Eddington-limited accretion  which lasts as long as the fuel supply
is available and eventually tapers off into periods of sub-Eddington
accretion.
  The duty cycle of the radio jet phase, associated with the
sub-Eddington accretion phase, 
is correspondingly small.

 The SMBH grows mostly in the super-Eddington phase while 
most of the spheroid stars  grow during the Eddington phase.  The latter 
phase ends by
quenching the feeding source, when the outflow clears out the remaining
gas. Only then is spheroid star formation terminated.  Hence the SMBH forms
before of order half of the spheroid stars.  

To clarify the connection
between the AGN outflow luminosity and star formation rate, I  write the
critical condition for gas outflow when the injected momentum flux, including
both sources, can no longer be contained by self-gravity.  One has
$$ M_g\dot v=L_{Edd}/c +\dot M_{outflow}v_\infty -GMM_g/r^2,$$ where I have
included the momentum input both from the Eddington-luminosity-limited
accreting and radiating SMBH and the supernova-driven outflows associated
with protogalactic star formation, and $v_\infty$ is the asymptotic wind flow
velocity.
Disruption occurs if $L_{Edd}/c+\dot M_{outflow}v_\infty$ exceeds
$(M_g/M)\sigma_g^4/G.$
As before, the outflow rate is related to the spheroid velocity
dispersion via the star formation rate, $\dot M_{outflow}=Qf_L\dot M_\ast.$

It follows that
$$M_{BH}=\frac{\kappa}{4\pi G^2}
\sigma^4\left(\frac{M_g}{M}\right)
\left(
1-f_L\alpha_S(\frac{\sigma_f}{\sigma_{g}})^{2.7}(v_w/\sigma_g)
\right).
$$
I infer that supernova-driven galactic outflows dominate until
$\sigma_{g}\approx 
(f_L \alpha_S v_w)^{0.3}
\sigma_f^{0.7},$
which   can be as  large as $\sim 100\rm
km \, s^{-1}.$ 
At larger gas turbulence velocities, black hole-driven
outflows dominate. This provides a possible demarcation in star formation efficiency
at  around $M_\ast \sim 3\times 
10^{10}\rm M_\odot $, resembling a trend seen in the SDSS data 
(Kauffmann et al. 2003).
if  BH outflow domination is associated with higher efficiency.

The main outcome is the observed relation between black
hole mass and spheroid velocity dispersion.
Moreover, the variance in the predicted
correlation between black
hole mass and spheroid velocity dispersion is likely to be controlled and kept small by the
self-regulation between supernova-stimulated  and Eddington wind-triggered star formation.
 If some  outflows are  sub-Eddington, as seems likely over long times,
an asymmetric variance results in the relation between black
hole mass and spheroid velocity dispersion. 

I now compare the two modes of star formation, writing $\dot
M_\ast=\alpha_{S,J}f_g\sigma_g^3/G,$ where $\alpha_S=\sigma_g
v_c/E_{SN} m_{SN}$ and $\alpha_J$ is set 
by the super-Eddington outflow,
$L_J\sim f_cv_hv_c^2\sigma_g^2/G.$
The  super-Eddington phase  is plausibly associated with  the quasar phenomenon, 
  the black hole outflow being  aided and abetted by triggered massive star formation.

The outflow is super-Eddington until the cocoon is limited by  ambient
pressure and becomes quasi-spherical.
As the massive star formation/death rate
slows, the AGN feeding augments and the outflow stimulates more star
formation.  The star formation has negative feedback on AGN feeding,
the AGN feeding has positive feedback on star formation.  If this
conjecture is correct, then 
$\dot M_\ast^E\sim L_{J}/cv_\infty \sim 
 f_c\sigma_g v_wv_c^2/v_JG,$
so that 
$\alpha_J \sim (f_c/f_g)(v_w/v_J) (v_c/ \sigma_g)^2  \sim 1.$

One could speculate that the super-Eddington phase 
is biased towards 
forming primarily
massive stars.  Such a top-heavy IMF is actually inferred in the
Arches cluster (Stolte et al. 2002), the most massive young
galactic star cluster
 and within 30
pc of the central supermassive black hole. One could even speculate that the
massive OB stars found within a parsec of SgrA$^{\ast}$ could have been formed by a brief phase of SMBH outflow
  a   million or so years ago.
 A top-heavy IMF would also
help explain the enrichment and abundance ratios in intracluster gas (Nagashima et al. 2005)
and in quasar emission line regions (Dietrich et al. 2003),
 the very high luminosities inferred in some ULIRGs
relative to the available gas supply,
and the SCUBA counts of submillimetre
galaxies (Baugh et al. 2005).
A top-heavy IMF  would also provide an attractive source of black hole
seeds for SMBH growth.
The high $[\alpha/Fe]$ ratios observed systematically in massive spheroidal
galaxies can be qualitatively  understood
since both $\alpha_S$ and  $\alpha_J$ are proportional
to $\sigma_g.$
Hence in lower  mass spheroids the low mass stars form relatively 
less  efficiently and hence more slowly, thereby diluting   the enhanced
$[\alpha/Fe]$,
 associated with massive star SNII yields, with the iron from 
low mass star SNIa yields.

There are other consequences of this model. The cocoon expansion velocity
 satisfies
$v_c=(v_hv_J)^{1/2}(R_h/R_c),$ so that in the 
 super-Eddington phase
 $\dot M_\ast\propto v_c^2v_w\sigma_g/v_J \simpropto v_w^2,$
since $v_h\approx v_c.$
Hence
the naive prediction of the model is that for ULIRGs, one should find
flows with $v_w\simpropto \dot M_\ast^{1/2}.$ Moreover these, in
contrast to supernova-driven flows which are driven by bubble energy
of the hot phase, are momentum-driven by the Compton-cooled AGN
wind. Neutral flows are expected, and the trend seen in NaI absorption
for ULIRG cold outflows is consistent with this model prediction
(Martin 2005). 

The Eddington phase coupling of wind and induced starburst-driven
outflows suggest that one will generally have $\dot
M_\ast=\alpha_S\sigma_g^3/G\propto \sigma_g^4,$ yielding an accounting
for the Faber-Jackson relation with approximately the correct
normalisation ({\it c.f.} Murray, Quataert and Thompson  2005).

Finally, I turn to core scalings.
Begelman and Nath (2005) argue that feedback energy is regulated by
the momentum deposition on the accreting gas in the region where the outflow
from the black hole   first undergoes strong cooling. The accretion is
Eddington-limited and this yields the correlation between core velocity
dispersion and black hole mass. The model presented here similarly has
momentum coupling and so not surprisingly  yields a similar scaling.
The final stellar core size is expected to be of order the Bondi radius,
within which black hole gravity dominates over that of the spheroid. 
However the detailed core properties, and in particular the
core scale,  must also depend on the extent  of the region where the
momentum is effectively injected into the accretion flow. The following
simple model illustrates the impact of feedback both on black hole growth and
on the core scale.

The onset of Compton cooling determines the end of the energy-conserving
phase of the jet and the onset of the momentum-driving that generates the
cocoon and the surrounding dense shell. I argue that this sets the core
scale, which coincidentally can be shown to be  of order the Bondi radius.  
The AGN
relativistic jet-induced plasma outflow undergoes Compton cooling at a radius
determined by setting the flow time equal to the Compton cooling time-scale.
This yields $R_c=M_{BH}\left(\frac{m_p}{m_e}\right)^2\frac {Gv_w}{c^3}, $
where $v_w\simgt  0.1 c.$ Comparing $R_c$ to the zone of influence of the SMBH,
the Bondi radius, $R_B=GM_{BH}\sigma_g^{-2},$ one finds that
$R_c=R_B\left(\frac{\sigma_g}{c}\right)^2
\left(\frac{m_p}{m_e}\right)^2\frac{v_w}{c}\sim R_B.$

Now $ R_B$ is a plausible scale for the cores of spheroids, as supermassive
black holes,
and in particular decays of  massive binary black holes (Lauer et al. 2005),
  are conjectured to have played a role in determining the core 
cuspiness or lack thereof by dynamical interactions with the spheroid stars.
One can now demonstrate that the core mass is $G^{-2}\beta \sigma_g^6,$ where
$\beta = {f_g} \frac{\sigma_T}{m_p}
\left(\frac{m_p}{m_e}\right)^2\frac{v_w}{c^3} \approx 10^{-16}f_g(v_w/c)\rm \,  s^2g^{-1}.$
This leads to two predictions.

The core phase space density (assuming the cooled gas forms the core stars) 
is 
$$\frac{\rho_c}{{\sigma_g}^3}=\beta^{1/2}f_g^{-1} M_c^{-5/2}
\propto M_c^{-5/2}. $$
This is close to the observed core scaling of core phase space density with mass
(Carlberg 1986, Faber et al. 1997).

Also the ratio of SMBH to core mass is
$ M_{BH}/ M_c=
\sigma_g^{-2}\left( \frac{m_e}{m_p}\right)^2
\frac{c^3}{v_w}\propto M_c^{-1/3}.$
This means that there is a greater reservoir of cooled gas for the more massive black holes.
Hence formation of massive SMBH should be favoured over those of lower mass. 
This provides at least a qualitative  anti-hierarchical explanation of the observed AGN x-ray luminosity function dependence on
redshift (Hasinger, Miyaji and Schmidt  2005). At the same time,
 the coupling of  SMBH outflow to spheroid star formation means that 
spheroid formation is equally anti-hierarchical, with massive spheroids in place before the lower mass systems.

\section{Conclusions}

Disk star formation is envisaged as a series
of ministarbursts. In a disk,  $\sigma_{g}\approx 10 \,\rm
km\, s^{-1}$, so that $\epsilon \approx 0.05$ for $\sigma_f\approx 
30 \,\rm
km\, s^{-1}.$
The observed efficiency is 
inferred from $\dot\Sigma_\ast = 0.017\Sigma_{gas}\Omega$, and is  globally about 2 percent.
This means that $\alpha_S\approx 0.02$ and 
$Q\sim 0.5,$ as is observed for the Milky Way.

In a starburst, however, the star formation efficiency is necessarily
higher because $\epsilon\propto \sigma_{gas}^{2.7}.$ The surprising 
phenomenological result is that starbursts also lie on the same
Kennicutt fit, $\dot\Sigma_\ast = Q\epsilon \Sigma_{gas}\Omega,$ with
$Q\epsilon=0.017.$ However there are two 
noteworthy differences with quiescent
disk star formation. The porosity is low since $\epsilon$ is
necessarily large. And the overall efficiency of star formation is
necessarily high  because of the enhanced stellar scale length  
associated with turbulence and ultimately responsible for driving the
starburst. Indeed, spatially extended star formation is seen in mergers. 
Self-regulation must account for the apparent conspiracy between $Q$ and $\epsilon.$

Starbursts are ubiquitous at high redshift.  In fact, starbursts are limited 
by the local gas supply.
Infalling satellites yield a stochastic gas supply.
A simple 
self-regulation hypothesis yields the star formation rate in a range of
 physical situations, thereby accounting for the Schmidt-Kennicutt law.
One can  regard quiescent disk star formation as a series of 
mini-starbursts, which seamlessly progresses into the major starburst regime
 as the gas supply rate  is enhanced. The fact that the
 phenomenological
star formation law accounts both for  quiescent disks and low mass starbursts can therefore be
 accommodated. 

The difference in star formation characteristics
 between massive and low mass disks has received attention
in a recent study of edge-on disks (Dalcanton, Yoachim and Bernstein 2004).
The case is made
that the scale height increases sharply (by a factor of about 2) as
the rotational velocity drops below 120 km/s, simultaneously with an
increase in disk gravitational stability to axisymmetric
perturbations.  This could be due to an increase in gas turbulence or
to a decrease in disk surface density.  
At a given disk mass, one would expect star formation
efficiency to globally decrease with higher turbulence,
due to the increased effects of disk venting via chimneys and fountains.
 This trend may 
be seen  in the edge-on disk sample, although detailed simulations
are needed in order to explore the theoretical implications that
involve the interaction of disk gravitational instabilities, molecular
cloud evolution, disk outflows and star formation.
One generally finds that lower mass disks tend to be  gravitationally stable 
and are not self-regulating.
The star formation rate is low presumably because the gas
supply provided by gravitational instability is reduced. Indeed, 
systematically lower star
formation efficiency in low rotation, low mass disks seems to be
indicated by the data. 

A plausible outcome of a starburst is that the mass in gas ejected is of
order the mass in stars formed.   For
low mass systems, the outflow should escape in a porosity-driven hot
 wind. It is less certain as to where the gas is
ejected.  It might be recycled into the disk via fountains and chimneys.

Massive galaxies require a more efficient driver to generate
ultraluminous starbursts and winds.  Supernovae do not provide enough
momentum input to drive winds from $L_\ast$ galaxies. For this,
recourse must be had to AGN-triggered outflows (Benson et al. 2004)
in order to inhibit gas accretion which otherwise results in
overproduction of overly luminous galaxies in the local 
galaxy luminosity function.
These
have the dual purpose of stimulating protogalactic star formation and
ultimately driving a strong galactic wind once the black hole mass
saturates at the Magorrian-Gebhardt-Ferrarese correlation.  Strong
winds lead in turn to saturation of the global star formation rate as
well as of the supermassive black hole mass. The starburst luminosity
is coupled to the Eddington luminosity, which is self-limiting with
regard to the gas supply.  The star formation rate and Eddington
luminosity are anti-correlated and self-regulated via feedback. 
The predicted AGN triggering of the extreme starbursts
associated with protospheroid formation means that AGN signatures should be
subdominant but still present in ultraluminous starbursts.

Black holes are in place before most of the spheroid stars have
 formed, at least in a massive spheroid, since it is the black hole
 outflows which provide the positive feedback stimulus for massive
 spheroid star formation.  
Observations of the $M_{BH}, \sigma_g$ correlation in AGN even 
 suggest that  SMBH formed before the full  spheroid potential 
was in place at very high redshift  (Walter et al.  2004).
However the situation is less clear  in the nearby universe:
SMBH masses determined by AGN emission
 line reverberation mapping
seem to lie both above
(Peterson et al. 2005; Treu, Malkan
 and Blandford 2004) and below (Grupe and Mathur 2004)
the ($M_{BH}, \sigma_\ast$) correlation, 
as do, respectively, spectroscopic  determinations  for an AGN (Silge et al. 2005)
and 
 a nearby SB0 galaxy (Coccato et al. 2005).

The supermassive black hole correlation is
 a natural outcome of gas accretion, SMBH growth and
 feedback-triggered star formation in the protospheroid.
One should also find examples of low mass SMBHs without spheroids
as the feedback argument must inevitably become ``leaky'' at low mass.
A brief super-Eddington phase is  associated with radio-loud quasars
and plausibly with top-heavy star formation, and can help account for rapid SMBH formation at very high redshift.
Compton-cooled flows  are another consequence of outflow-triggered star formation, in that a dense gas reservoir is provided in the
cores for SMBH and spheroid star formation. The inferred scaling laws for core phase space density match the observed trend.
Anti-hierarchical SMBH formation is a natural consequence of massive core formation,
and this also means that the stellar cores will follow a similar trend with spheroid mass.


\section*{acknowledgments}

I am grateful to the Kapteyn Astronomical Institute at Groningen and
to the Director Prof. Piet van der Kruit,  
 to KITP and the organisers of the
Galaxy-IGM Interactions Program, and to IAP, Paris for hospitality when I was
preparing this paper, and to Tel Aviv University for appointment as a Sackler Scholar,
where this work
 was completed.
\def\mnras{MNRAS}
\def\araa{ARAA}
\def\apj{ApJ}
\def\aj{AJ}
\def\pasp{PASP}
\def\apjl{ApJ}


\end{document}